\documentclass[aps,twocolumn,amssymb,amsfonts,amsmath]{revtex4-1}
\usepackage{amssymb}
\usepackage{graphicx}
\usepackage{dcolumn}
\usepackage{color}
\usepackage{caption}

\begin{document}
%\linenumbers

%%%%%%%%%%%%%%%%%%%%%
% Title and authors %
%%%%%%%%%%%%%%%%%%%%%

\title{Effect of exchange-correlation spin-torque on spin dynamics}

\author{J. K. Dewhurst}
\author{A. Sanna}
\author{S. Sharma}
\affiliation{Max Planck Institute of Microstructure Physics, Weinberg 2, D-06120 Halle, Germany}
\email{sharma@mpi-halle.mpg.de}

\date{\today}

%%%%%%%%%%%%
% Abstract %
%%%%%%%%%%%%

\begin{abstract}
A recently proposed exchange-correlation functional with in density functional theory, which ensures that the exchange-correlation magnetic field is source-free, is shown to give non-zero internal spin-torque. This spin-torque is  identically zero for all conventional local and semi-local functionals. Extension of this  source-free functional to the time domain is used to study the effect of the internal spin-torque on the laser induced spin-dynamics in bulk Co, Ni and interfaces of these metals with Pt. It is shown that the internal spin-torque contribute significantly to spin-dynamics only when the magneto crystalline anisotropy energy is small, as in the case of cubic bulk materials. For surfaces or interfaces, where the anisotropy energy is large, these torques are too small to cause any significant precession of spins in early times ($<$ 100fs). Further more it is shown that the spin-dynamics caused by  the internal spin-torque is slow compared to the inter-site spin transfer and spin-orbit mediated spin-flips. 
\end{abstract}

\maketitle

%%%%%%%%%%%%%%%%%%%%%%
% Main body of paper %
%%%%%%%%%%%%%%%%%%%%%%

\section{Introduction}
The possibility of controlling electronic spins by light offers a future of highly efficient devices and fast (sub femtosecond) memory storage. In light of this a large amount of research is being devoted to the study of laser induced dynamics of spins-- spin-injection\cite{LBD05,CHWA09,RLBA12,MRWP11,HMSB17}, spin transfer torque\cite{NRTT12,SKWK14,RAIM17,LHK17} across tailored interfaces, all-optical switching\cite{RVSK11,kimmel,oistr}, ultra-fast demagnetization\cite{BMDB96,SBJE97,HMKB97,ABPW97,KKR10,KKKJ00,MWDM09,CDFG12,SAMC12,SPSG14,ESMB14,KSKP11,KKTP04,FDSV08} to name but a few examples. 

Theoretically, \emph{ab-initio} methods for treating this laser-induced spin-dynamics is the non-collinear spin-polarized extension time-dependent density functional theory (TD-DFT). The requirement of non-collinearity stems from the fact that, to leading order, light couples to spins via the spin-orbit (SO) coupling term, which requires that the Kohn-Sham wave-functions be two component spinors.  In principle, TD-DFT is an exact method, but in practice the  quality of results depends upon the approximation used for exchange-correlation (xc) energy functional. 

Usually, in time-dependent case, the adiabatic extension of ground state xc functionals is used for time-propagating the Kohn-Sham system. Most of the xc functionals like the local spin density approximation (LSDA) or generalized gradient approximation(GGA) are designed for collinear systems and a non-collinear extension of these functionals is performed using the Kubler-Sandratskii method\cite{kubler88,sandratskii86}-- at each point in space and time the densities (charge density $\rho$, magnetization density {\bf m}), which are $2 \times 2$ complex matrices in spin-space, are first diagonalized and then the corresponding xc potentials ($v_{\rm xc}$, {\bf B}$_{\rm xc}$) are calculated via functional derivatives of energy wrt these diagonal densities. This immediately implies that by construction {\bf m} and {\bf B}$_{\rm xc}$ are parallel at each point in space and time and torque ${\bf m}({\bf r},t) \times {\bf B}_{\rm xc}({\bf r},t) = 0$, which is equivalent to saying that the internal torque felt by the spins is identically 0. This is a serious limitation as once the external perturbation (like magnetic field or laser) have been switched off, it is these torques which contribute to the dynamics of the spins. 
This raises an interesting question: if one were to design a truly non-collinear functional, like the optimized effective potential\cite{our-exx}, which gives a non-zero torque on the spins, would the laser induced spin-dynamics be fundamentally different from one observed for conventional functionals like adiabatic LSDA\cite{lda}?

In order to answer this question, in the present work we employ time-dependent extension of our recently developed source-free functional\cite{srf} to study the laser-induced spin-dynamics. The source-free functional is a truly non-collinear functional and, as we demonstrate in the present work, it leads to a non-zero torque on the spins. We find that for bulk systems (Ni and Co), where SO induced anisotropy is very small i.e. magneto crystalline anisotropy (MCA) energy is only 2 $\mu$eV/atom, internal torques on spins lead to precession of spins about the easy axis an effect which cannot be described by conventional functionals like ALSDA. For surfaces and interfaces, where the MCA is $\sim$1meV/atom, internal torques do not cause much precession of spins and results for ALSDA and source-free functional are almost the same.

\section{Methodology}
\subsection{Time-dependent density functional theory}
The Runge-Gross theorem \cite{RG84} establishes that the time-dependent external potential is a unique functional of the time dependent density, given the initial state. Based on this theorem, a system of non-interacting particles can be chosen such that the density of this non-interacting system is equal to that of the interacting system for all times\cite{EFB09,C11,SDG14}. The wave function of this non-interacting system is represented as a Slater determinant of single-particle orbitals. In what follows a fully non-collinear spin-dependent version of these theorems is employed\cite{KDES15}. Then the time-dependent Kohn-Sham (KS) orbitals are 2-component Pauli spinors, $\psi$, determined by the equations:

\begin{eqnarray}\label{KS}
i\frac{\partial \psi_j({\bf r},t)}{\partial t} &=& \left[
\frac{1}{2}\left(-i{\nabla} +\frac{1}{c}{\bf A}_{\rm ext}(t)\right)^2 +v_{s}({\bf r},t)  \right. \\ \nonumber
&+& \left.\frac{1}{2c} {\sigma}\cdot{\bf B}_{s}({\bf r},t) + \frac{1}{4c^2} {\sigma}\cdot ({\nabla}v_{s}({\bf r},t) \times -i{\nabla})\right]
\psi_j({\bf r},t)
\end{eqnarray}
where ${\bf A}_{\rm ext}(t)$ is a vector potential representing the applied laser field, and ${\sigma}$ are the
Pauli matrices. The KS effective potential $v_{s}({\bf r},t) = v_{\rm ext}({\bf r},t)+v_{\rm H}({\bf r},t)+v_{\rm xc}({\bf r},t)$ is decomposed into the external potential $v_{\rm ext}$, the classical electrostatic Hartree potential $v_{\rm H}$ and the xc potential $v_{\rm xc}$. Similarly the KS magnetic field is written as ${\bf B}_{s}({\bf r},t)={\bf B}_{\rm ext}(t)+{\bf B}_{\rm xc}({\bf r},t)$ where ${\bf B}_{\rm ext}(t)$ is the magnetic field of the applied laser pulse plus possibly an additional magnetic field and ${\bf B}_{\rm xc}({\bf r},t)$ is the xc magnetic field. The final term of Eq.~\eqref{KS} is the spin-orbit coupling term. It is assumed that the wavelength of the applied laser is much greater than the size of a unit cell and the dipole approximation can be used i.e. the spatial dependence of the vector potential is disregarded.  The 2-component Pauli spinors, $\psi$, are then used to construct the magnetization density as:
\begin{equation}
{\bf m}({\bf r},t)= \sum^N_{j=1} \psi^{\dagger}_j({\bf r},t) \sigma \psi_j({\bf r},t),
\label{mr}
\end{equation}
making {\bf m} a $2 \times 2$ matrix in the spin-space.

\subsection{Functional}
In order to  propagate Eq.~(\ref{KS}) in time one needs to approximate $v_{\rm xc}$ and ${\bf B}_{\rm xc}$. Usually adiabatic extensions of ground-state functionals like LSDA are used in such time-propagation scheme\cite{yabana,KDES15,SSS17,KEMS17,SSBM17}. LSDA is a collinear functional by design and requires only $\rho^{\uparrow}$ and $\rho^{\downarrow}$ as input. Thus only $v_{\rm xc}$ and ${\bf B}^{z}_{\rm xc}$ are obtained from the functional derivative. A non-collinear extension can be constructed by first diagonalizing the $2 \times 2$ magnetization density in Eq. (\ref{mr}) at each point in space and then calculating ${\bf B}_{\rm xc}$ by taking functional derivative of energy with respect to this diagonal magnetization density and reversing the diagonalization . Such a ${\bf B}_{\rm xc}$ is not curl of a vector field and contains unphysical magnetic monopoles (i.e. source terms). More importantly, by construction such a functional gives ${\bf m}({\bf r},t) \times {\bf B}_{\rm xc}({\bf r},t) = 0$ i.e. the internal torque on spins is zero at all times and in all space. 

Recently, it was shown that when these unphysical source-terms are removed from ALSDA\cite{srf} the resulting functional is a truly non-collinear functional. This functional was able to successfully describe the correct ground-state moment of pnictides, a problem which has been intractable for well over a decade. Interestingly, being a truly non-collinear functional the source-free ALSDA gives torque ${\bf m}({\bf r},t) \times {\bf B}_{\rm xc}({\bf r},t) \ne 0$ and hence would contribute to the dynamics of the spins. The construction of this source-free functional requires following 3 steps:
\begin{itemize}
\item The LDA energy functional is modified as $E_{\rm xc}[\rho,{\bf m}]\rightarrow E_{\rm xc}[\rho,s{\bf m}]$ and a scaling of the resultant xc field is performed as ${\bf B}^{\rm LDA}_{\rm xc}\rightarrow s{\bf B}^{\rm LDA}_{\rm xc}$ in order to keep the functional variational with respect to ${\bf m}$. The value of $s$ is chosen empirically to be  1.12.
\item Following Poisons equation is then solved to calculate $\phi$ 
\begin{align}
 \nabla^2\phi({\bf r},t)=-4\pi\nabla\cdot{\bf B}^{\rm LDA}_{\rm xc}({\bf r},t).
\end{align}
\item Finally the source term is removed from  ${\bf B}^{\rm LDA}_{\rm xc}$ using:
\begin{align}
 {\bf B}^{\rm SF}_{\rm xc}({\bf r},t)\equiv{\bf B}^{\rm LDA}_{\rm xc}({\bf r},t)
 +\frac{1}{4\pi}\nabla\phi({\bf r},t).
\end{align}
\end{itemize}
It is easy to show that $\nabla \cdot {\bf B}^{\rm SF}_{\rm xc} = 0$.

\section{Computational details}
All the calculations in the present work are done using the state-of-the art full potential linearized augmented plane wave (LAPW) method as implemented within the Elk code\cite{elk} code. Within this method the core electrons (with eigenvalues 95~eV below Fermi energy) are treated fully relativistically by solving the radial Dirac equation while higher lying electrons are treated using the scalar relativistic Hamiltonian in the presence of the spin-orbit coupling. To obtain the 2-component Pauli spinor states, the Hamiltonian containing only the scalar potential is diagonalized in the LAPW basis: this is the first variational step. The scalar states thus obtained are then used as a basis to set up a second-variational Hamiltonian with spinor degrees of freedom \cite{singh}. This is more efficient than simply using spinor LAPW functions, however care must be taken to ensure that a sufficient number of first-variational eigenstates for convergence of the second-variational problem are used.

We solve Eq.~(\ref{KS}) for the electronic system alone. Coupling of the electronic system to the nuclear degrees of freedom is not included in the present work. Radiative effects, which can be included by simultaneously time-propagating Maxwell's equations, are also not included in the present work. At longer times scales these effects are expected to contribute significantly.

A regular mesh in {\bf k}-space of $8\times8\times1$ for multi-layers and $8\times8\times8$ for bulk was used and a time step of $\Delta t=4.13$fs was employed for the time-propagation algorithm\cite{DKSG16}. A smearing width of 0.027~eV was used. Laser pulses used in the present work are linearly polarized with a frequency of 1.55~eV (red). For all ground-state calculations a full structural optimization was performed. For the case of Co/Pt(001) and Ni/Pt(001), the Pt substrate was simulated by using 4 to 8 Pt mono layers (ML). We found that for Pt layer thickness greater than 4 ML the results do not change significantly and hence all results presented here are for 3ML of Co or Ni on 5ML of Pt(001).

\section{Results}
\subsection{Ground-state and internal spin-torque}
The first step is to determine the ground-state of the material by using the LSDA and source-free functional. In order to ensure an unbiased magnetic ground-state a fully  unconstrained minimization was performed without imposing any magnetic symmetries. The moments thus obtained are shown in Table I. Then the laser pulse  is applied to the material and the evolution of spins are studied as a function of time using TD-DFT.  It is clear from these results that the ground-state obtained using the source-free functional is slightly more non-collinear than the LSDA.
\begin{table}[tbh] 
\caption{Ground-state moments (in $\mu_{\rm B}$) for all the materials studied in the present work. The results are calculated using LSDA and source-free functionals.}
\begin{tabular}{ c || c c c | c c c}
Material &     & LSDA    &       &    & Source-free & \\ 
         & M$_x$ & M$_y$ & M$_z$ & M$_x$ & M$_y$ & M$_z$ \\ \hline \hline
Bulk Ni  &   0   &  0    &  0.67 & 0.02  & 0.37  & 0.56  \\ \hline 
Bulk Co  &   0   &  0    &  1.69 &  0    & 0.11  & 1.69  \\ \hline
Ni@Ni/Pt &   0   &  0.78 &  0    & 0.04  & 0.76  & 0 \\  
         &   0   &  0.71 &  0    & 0.04  & 0.67  & 0 \\
         &  0.01 &  0.57 &  0    & 0.03  & 0.52  & 0 \\ \hline
Co@Co/Pt &  0    &  1.89 &  0    & 0     & 1.87  & 0 \\
         &  0    &  1.72 &  0    & 0     & 1.67  & 0 \\
         &  0.11 &  1.66 &  0    & 0.1   & 1.62  & 0 \\ \hline
\end{tabular}
\end{table}

\begin{figure}[h]
\centerline{\begin{tabular}{c}
\includegraphics[width=\columnwidth]{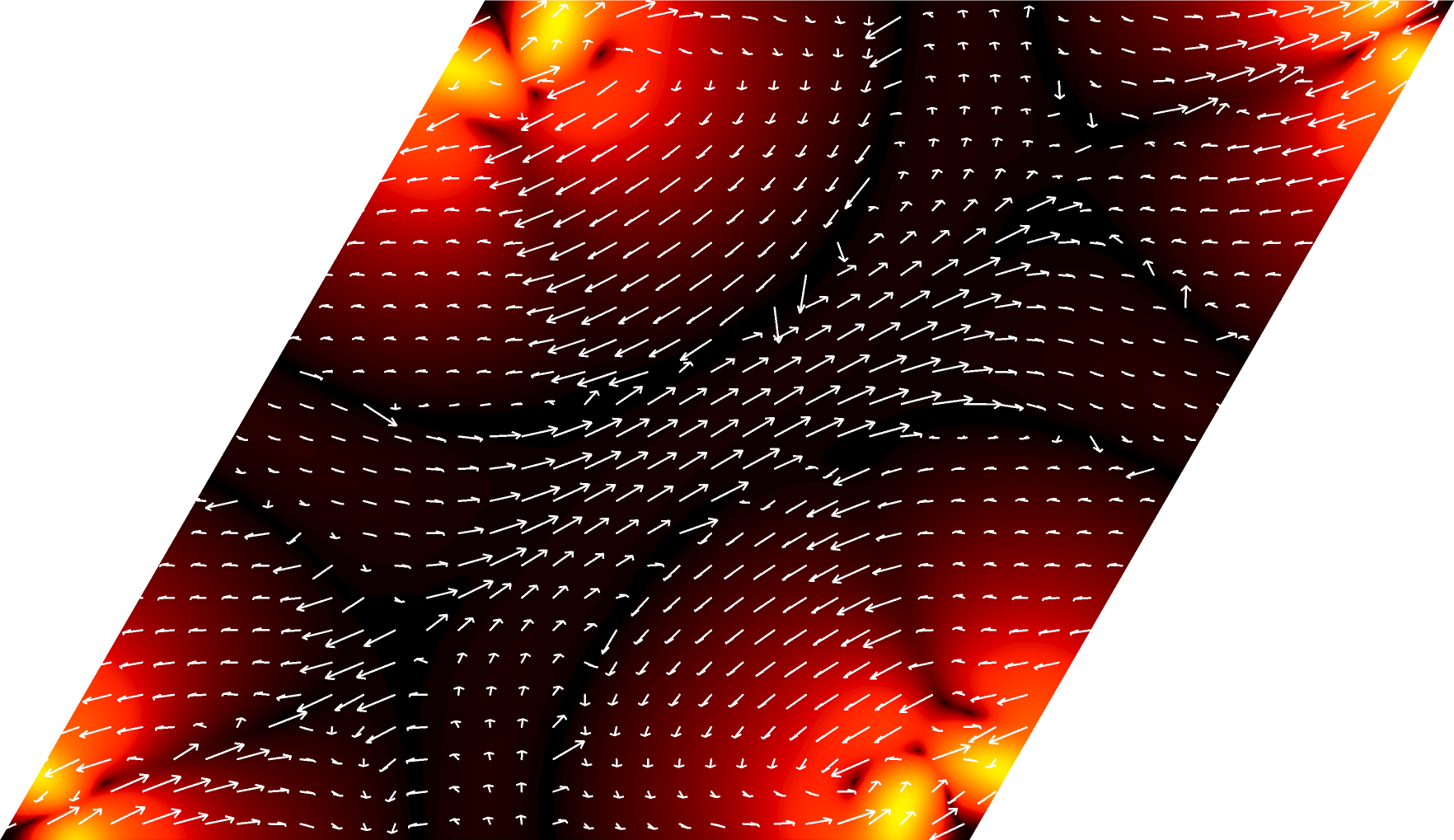} \\
 \\
\\
\includegraphics[width=0.4\columnwidth]{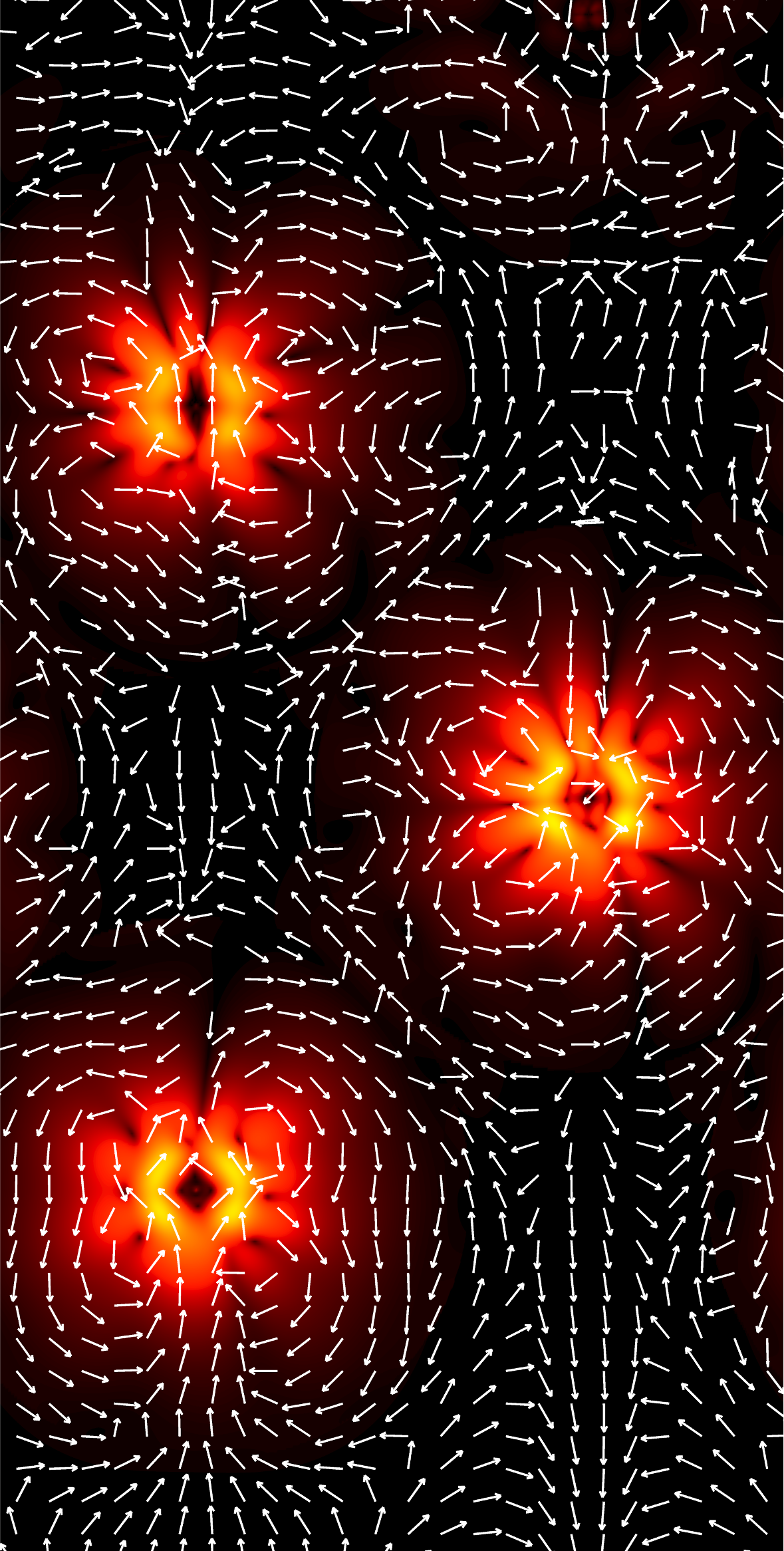}
\end{tabular}}
\caption{Top panel shows the ${\bf m}({\bf r},t=0) \times {\bf B}_{\rm xc}({\bf r},t=0)$ for bulk Ni in (111) plane. Bottom panel shows the same for 3Ni/5Pt in the (110) plane. The arows indicate the direction and colors the magnitude.}
\label{mcbxc}
\end{figure}
Unlike LSDA, source-free functional does not require  ${\bf B}_{\rm xc}$ to be parallel to {\bf m}. This results in the internal spin-torque, ${\bf m}({\bf r}) \times {\bf B}_{\rm xc}({\bf r})$, being non-zero. These are plotted these spin-torques are plotted  in Fig. \ref{mcbxc} for bulk Ni and a 3Ni/5Pt interface. In the ground-state, ${\bf m}({\bf r}) \times {\bf B}_{\rm xc}({\bf r})$ exactly cancels the divergence of spin-current\cite{our-exx}. However, this term can contribute to the spin-dynamics away from the equilibrium. In the next section we look at the effect of this internal spin-torque on the laser-induced dynamics of spins.

\subsection{Laser induced spin-dynamics}  
\begin{figure}[h]
\centerline{\begin{tabular}{c}
\includegraphics[width=\columnwidth]{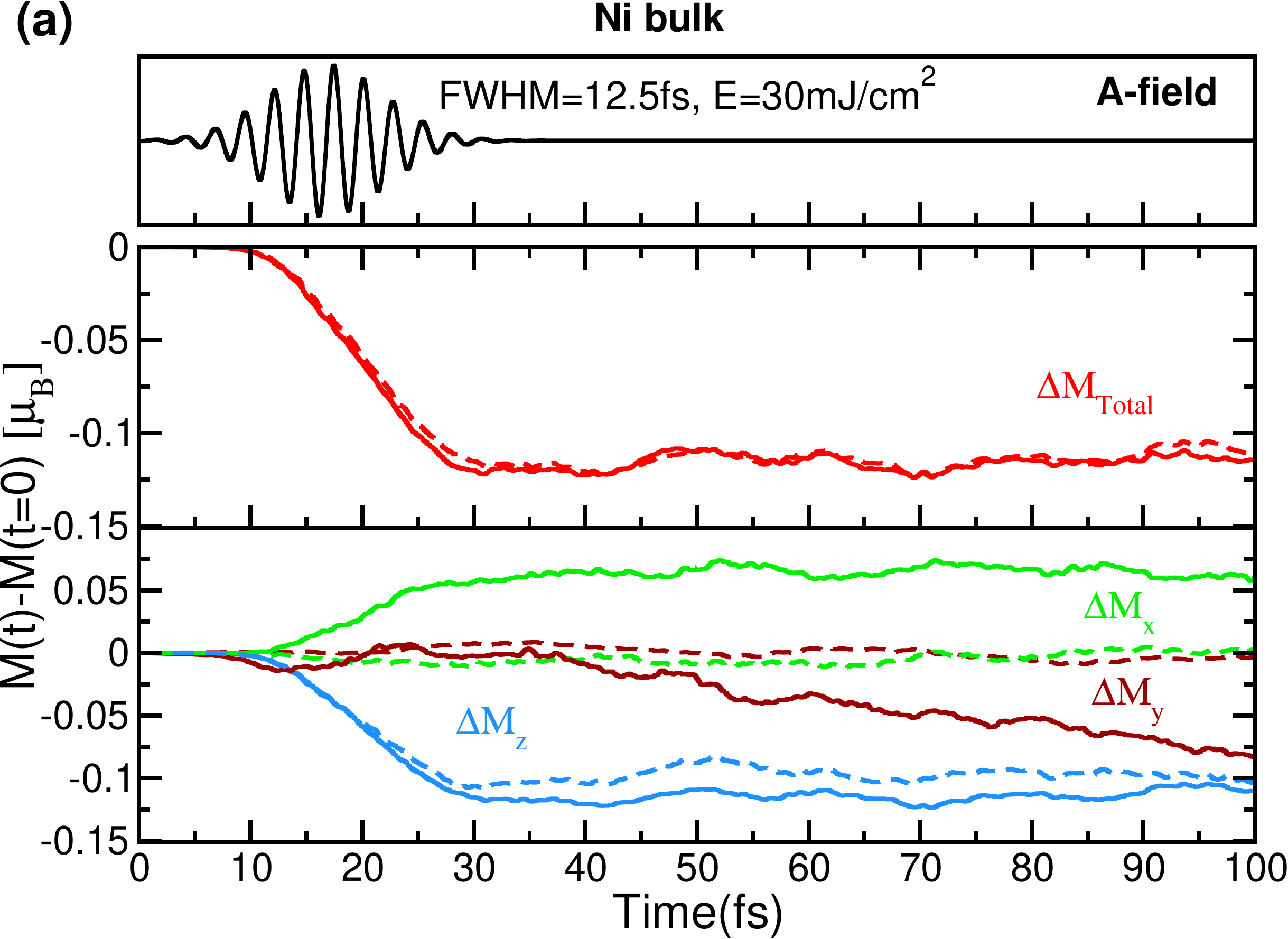} \\
\includegraphics[width=\columnwidth]{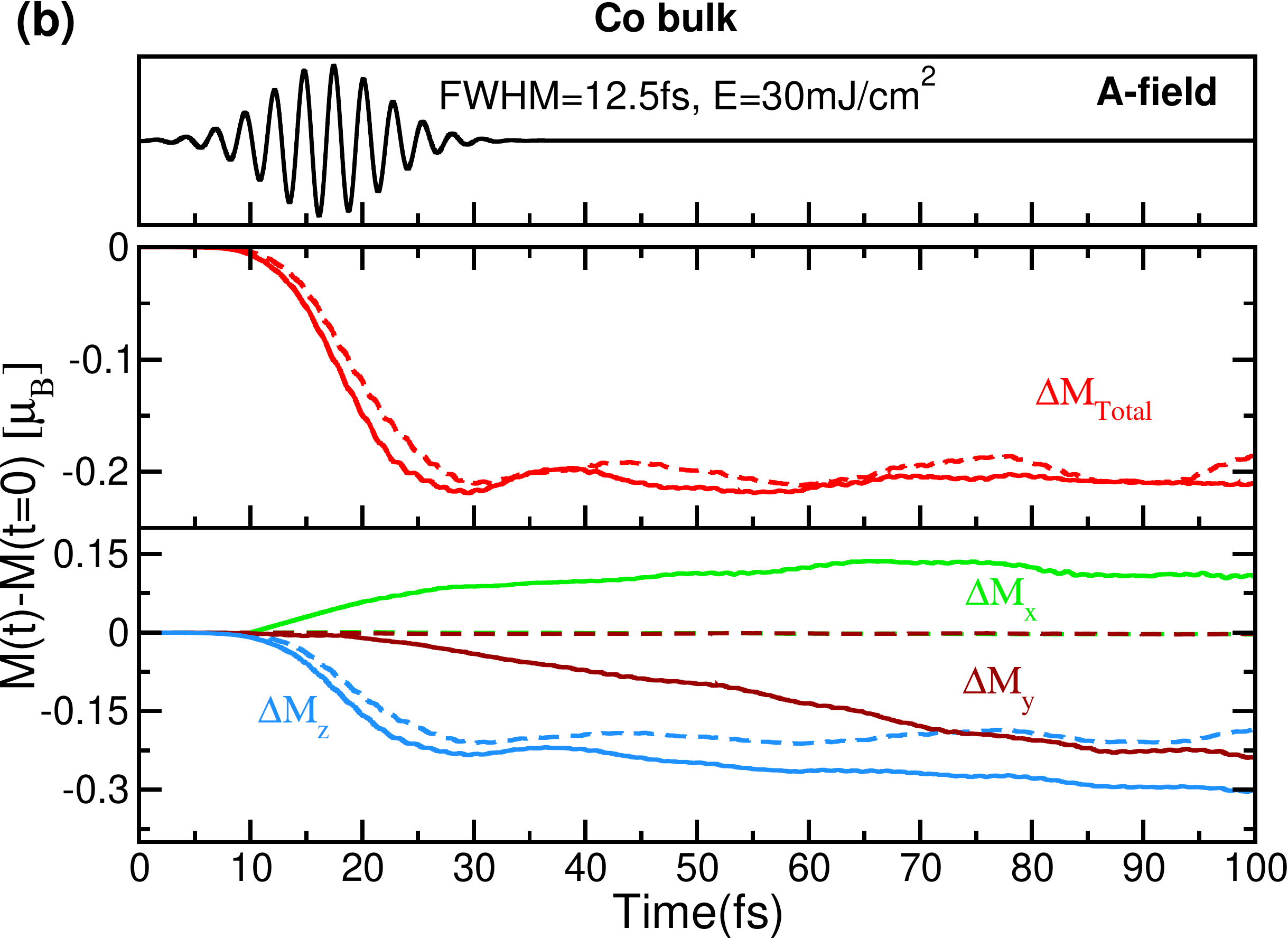}
\end{tabular}}
\caption{(a) Top panel shows the vector potential of the applied linearly polarized (along $z$-axis) laser pulse with frequency=1.5eV, FWHM=12.5fs and fluence = 30mJ/cm$^2$. Middle panel shows the total moment (red) and the bottom panel $x$ (green), $y$ (brown) and $z$ (blue) projected moments for bulk Ni as a function of time (in fs). Dashed line are the results obtained using ALSDA and full lines the results obtained using source-free functional. (b) The same as (a) but for bulk Co.}
\label{sd-bulk}
\end{figure}
We now focus our attention on laser induced spin-dynamics. The results for bulk Ni and Co are shown in Fig. \ref{sd-bulk}. The dynamics of the total moment, $M_{\rm total}=\sqrt{M_{x}^2+M_{y}^2+M_{z}^2}$ (shown in red), obtained using ALSDA and  source-free functional are almost the same. A closer look however reveals that the $x$, $y$ and $z$ projected moments are strikingly different for the two functionals. 
In the case of ALSDA there is no change in $M_{\rm x}$ and $M_{\rm y}$ as a function of time. On the other hand, internal torques in the source-free functional cause the spins to rotate around the $z$-axis as a function of time-- as $M_{\rm x}(t)$ increases, $M_{\rm y}(t)$ decreases. $M_{\rm z}(t)$ shows demagnetization for both the functionals the reason for which, we find to be, spin-orbit induce spin-flips. It is also important to note that during the first $\sim$30fs (when the laser pulse creates a non-equilibrium charge distribution) the effect of the internal spin-torque is small and all the spin-dynamics is caused by the SO induced spin-flips. This indicates that the contribution of the torque term to the spin-dynamics is slower than the SO term.

\begin{figure}[h]
\centerline{\begin{tabular}{c}
\includegraphics[width=\columnwidth]{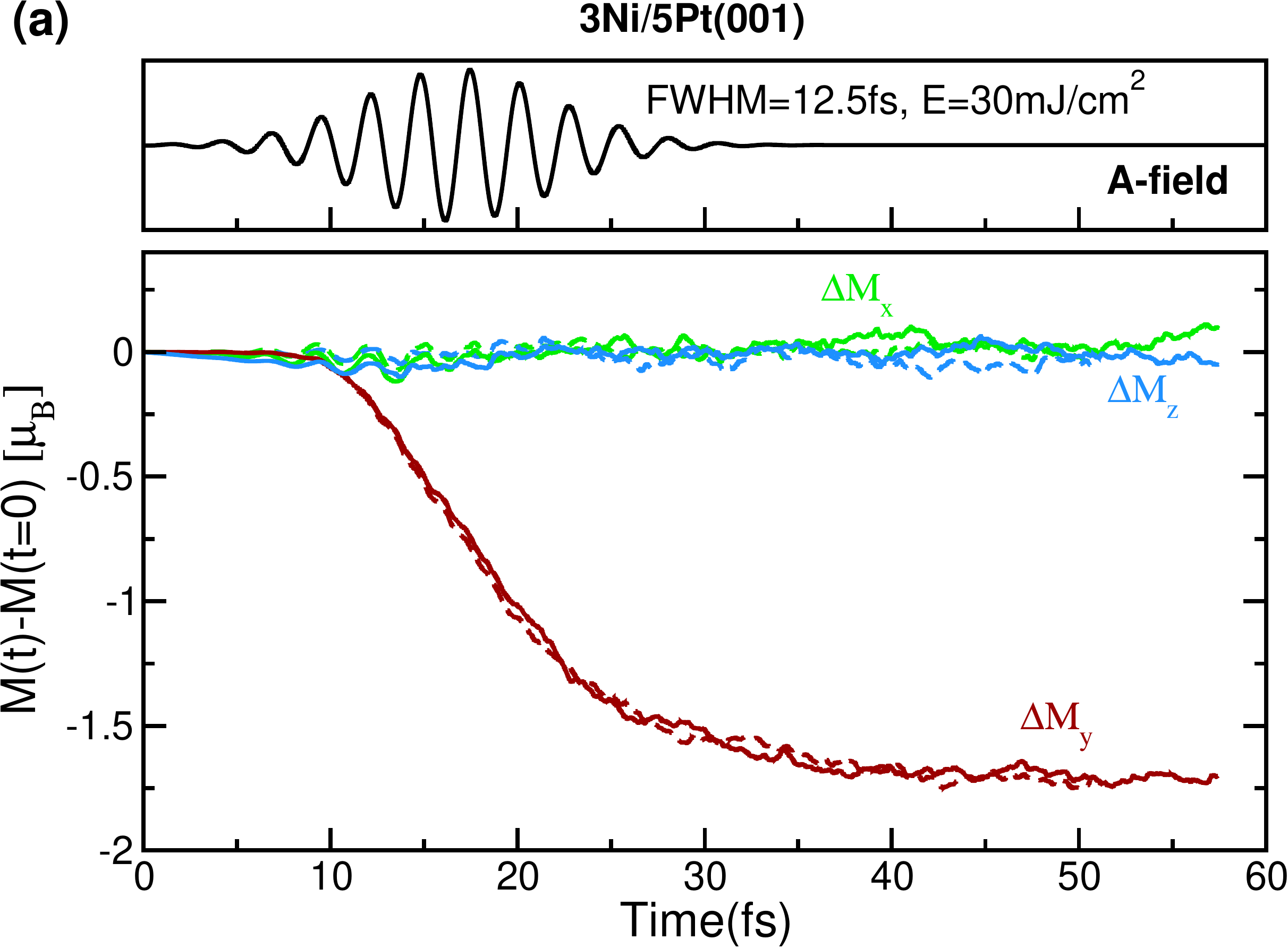} \\
\includegraphics[width=\columnwidth]{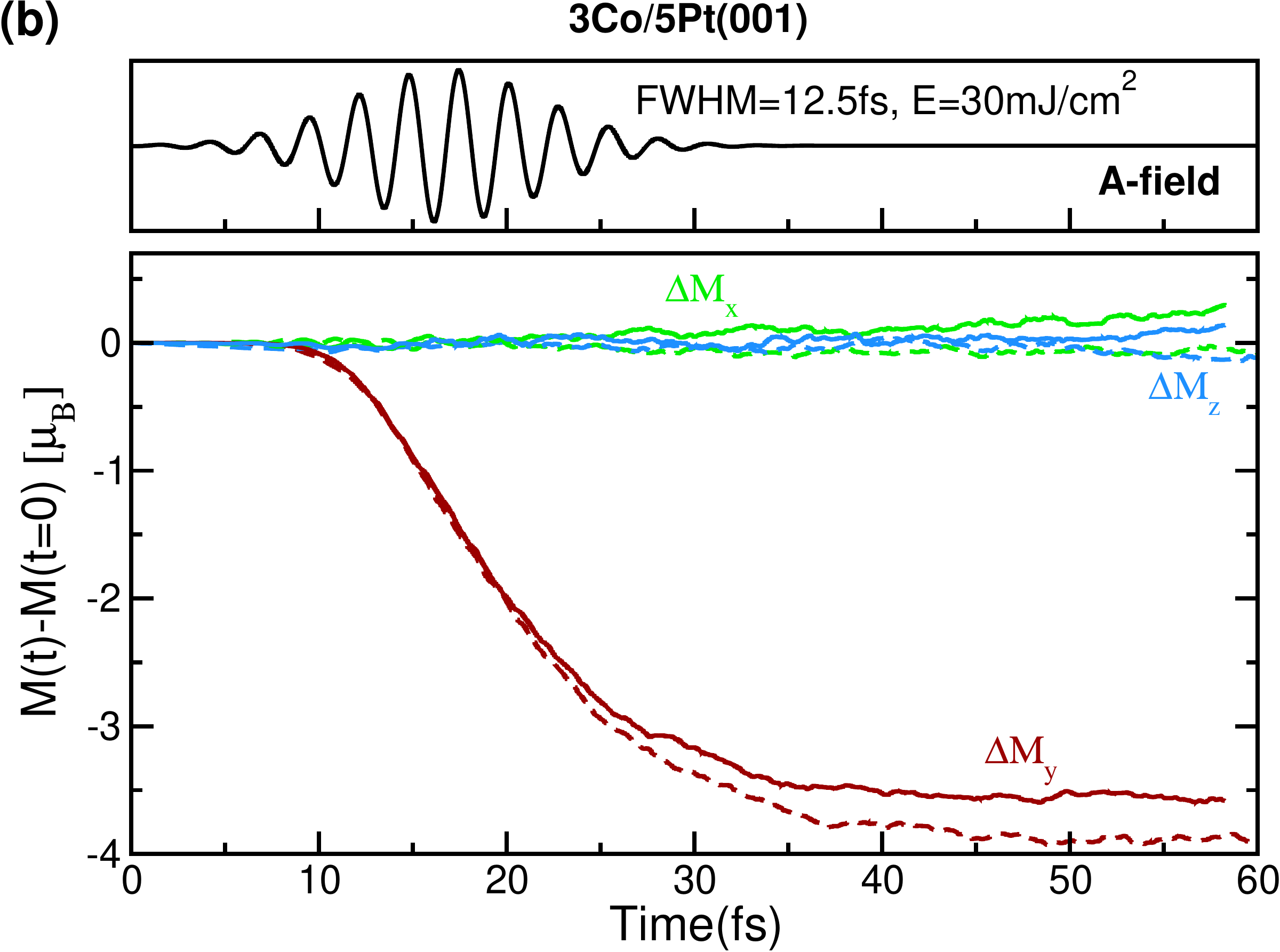}
\end{tabular}}
\caption{(a) Top panel shows the vector potential of the applied linearly polarized (along $y$-axis) laser pulse with frequency=1.5eV, FWHM=12.5fs and fluence = 30mJ/cm$^2$. Lower panel shows $x$ (green), $y$ (brown) and $z$ (blue) projected moments for 3ML of Ni on 3ML of Pt(001) as a function of time (in fs). Dashed line are the results obtained using ALSDA and full lines the results obtained using source-free ALSDA. (b) The same as (a) but for 3ML of Co on 5ML of Pt(001).}
\label{sd-inter}
\end{figure}
Since this internal spin-torque increases the non-collinearity in the system, it will be interesting to know how they effect the spin-dynamics on surfaces and interfaces, where lower symmetry  has the effect of frustrating and thus enhancing the non-collinear nature of spins. Such results are shown in Fig. \ref{sd-inter} for 3ML of Co on 5ML of Pt(001) and 3ML  Ni on 5ML of Pt(001). In both cases the ground-state moment points in-plane (see Table I). The laser induced spin-dynamics from ALSDA as well as source-free functional show that the internal spin-torque does not significantly contribute to spin-dynamics in early times ($<$100fs). In the case of interfaces the demagnetization of Ni layers is caused  by two distinct processes-- (i) spin injection in the Pt layers: optically excited electrons make a inter-site spin transfer. This effect is called OISTR\cite{oistr} and is caused due to optical charge excitations. These excitations  lead to majority spin electrons being injected into the Pt layers. OISTR dominates the physics of demagnetization for the first 25fs. (ii) Spin-orbit induced spin-flips\cite{KDES15,KEMS17,SSBM17}: this process dominates after the first 25fs.   

At the first sight these results look surprising -- the internal torque-induced spin-dynamics in bulk is much larger than for surfaces or interfaces. However, this can be explained based on ground-state energetics -- in the case of bulk Ni and Co the MCA ($\Delta E= E_{z}-E_{xy}$, where $E_{i}$ is the total energy with spins pointing in the $i$-direction) is or the order of .. $\mu$eV, making it easy for a small spin-torque (in Fig. \ref{mcbxc}) to rotate the electronic spin about the easy axis. In the case of interfaces the MCA is much larger, $\sim$1meV/atom, and small torque terms are not sufficient to overcome the energy barrier to rotate the spins significantly. 

\section{Summary}
In the present work we explore the effect of using recently derived source-free exchange-correlation functional on the ground-state and laser induced spin-dynamics of bulk Ni, Co and interfaces of these metals with Pt. We compare the results obtained using source-free functional to those obtained using unmodified LSDA. Our key findings are: 
(a) the source-free functional introduces an extra non-collinearity in the ground-state compared to the conventional LSDA functional, 
(b) the spin-torque term, which is identically 0 for LSDA, is non-zero for source-free functional, 
(c) the effect of this internal spin-torque on laser induced spin-dynamics is most prominent for systems where the magneto crystalline aninsotropy is small like bulk cubic materials. For such materials the small torque generated by the source-free functional are sufficient to cause spin precession about the easy axis and 
(d) the contribution of this spin rotation due to internal spin-torque is a slow process compared to optical inter-site spin transfer and spin-orbit induced spin-flips. 

In future it would be interesting to explore properties which are profoundly affected by spin-torque such as  magnon response, spin-transfer-torque and spin-orbit torque. 

\subsection{Acknowledgments}
Sharma would like to acknowledge SPP-QUTIF and SFB 762 for funding. 

%\bibliography{switch}

\end{document}